# Growth of high-mobility Bi$_2$Te$_2$Se nanoplatelets on BN sheets by van der Waals epitaxy

*Pascal Gehring*[*,†], *Bo F. Gao*[†,§], *Marko Burghard*[†], *Klaus Kern*[†,‡]

[†]Max Planck Institute for Solid State Research, Heisenbergstrasse 1, D-70569 Stuttgart, Germany

[‡]Institute de Physique de la Matière Condensée, Ecole Polytechnique de Lausanne, 1015 Lausanne, Switzerland

ABSTRACT

The electrical detection of the surface states of topological insulators is strongly impeded by the interference of bulk conduction, which commonly arises due to pronounced doping associated with the formation of lattice vacancies. As exemplified by the topological insulator Bi$_2$Te$_2$Se, we show that via van der Waals epitaxial growth on thin BN substrates, the structural quality of such nanoplatelets can be substantially improved. The surface state carrier mobility of nanoplatelets on hBN is increased by a factor of about 3 compared to platelets on conventional Si/SiO$_x$ substrates, which enables the observation of well-developed Shubnikov-de Haas oscillations. We furthermore demonstrate the possibility to effectively tune the Fermi level position in the films with the aid of a back gate.

Since their theoretical prediction in 2005[1], topological insulators (TIs) have been the subject of numerous theoretical and experimental studies. This extraordinary interest is driven by the fact that TIs are insulators in the bulk, but manifest conducting helical states at their boundaries[1,2,3]. In contrast to the quantum Hall state, these boundary states are the result of strong spin-orbit coupling without any external magnetic field, such that time reversal symmetry is conserved and accordingly backscattering is forbidden. The unique combination of dissipation-less charge transport channels with intrinsic spin polarization renders TIs of strong interest for spintronic applications, including spin injection[4,5] or spin transfer torques[6]. In addition, they hold promise for novel and exciting fundamental effects, like the formation of magnetic monopoles[7] or Majorana fermions[8].

While theory predicts TIs to be perfect bulk insulators, in practice their charge transport properties are dominated by defects. In fact, a sizeable amount of point defects (vacancies or anti-site defects) is commonly introduced during the crystal growth in case of the chalcogenides $Bi_2Se_3$, $Bi_2Te_3$, $Sb_2Te_3$ and their alloys, which represent the currently most intensively studied three-dimensional (3D) TIs. The presence of such defects leads to strong bulk doping in these compounds[3,9,10]. Moreover, chemical reactions at the crystal surface have been documented to impart notable doping[11,12]. The bulk doping creates a quasi-metallic conduction channel parallel to the surface channels, rendering it difficult to unequivocally assign charge transport features to the topological protected surface states. Strategies that have been pursued to minimize the contribution of the bulk transport include compensation doping[13,14], alloying of differently doped TIs[15,16], or increasing the surface to bulk ratio by using ultra thin samples[17]. However, doping or alloying typically results in a concomitant drastic decrease in carrier mobility due to the introduction of defects. One possibility to avoid such decrease in the structural quality of the crystals involves the epitaxial growth of TI thin films on graphene, which has been demonstrated to significantly reduce the defect density and hence the bulk interference[18]. At the same time, the lowered defect concentration in epitaxially grown films ensures enhanced carrier mobilities[19]. The suitability of graphene as a substrate for epitaxial growth of TI materials derives from its layered structure with hexagonal lattice symmetry, combined with its C-C bonding length of ~1.42 Å corresponding to a small lattice mismatch of only ~2.9% vs. $Bi_2Se_3$[18,25]. However, due to its high electrical conductivity, graphene rules out as underlying substrate if electrical transport studies are desired on the grown TIs.

Here, we report the van der Waals epitaxial growth and electrical characterization of high quality $Bi_2Te_2Se$ thin films on electrically insulating boron nitride sheets (band gap of about 6 eV[26]). $Bi_2Te_2Se$ has emerged as one of the most promising TIs due to its simple surface band structure, large bulk band gap and low bulk contribution to the total charge transport[16,20,21]. Our $Bi_2Te_2Se$ films grown by van der Waals epitaxy are sufficiently ordered to display Shubnikov-de Haas (SdH) oscillations which provide access to important quantities such as the cyclotron mass, the Fermi velocity or Fermi energy[23]. This capability is in contrast to previously investigated $Bi_2Te_2Se$ films grown by chemical vapor deposition (CVD) on $Si/SiO_x$ substrates, in which case the relatively low surface state mobility has prevented the observation of SdH oscillations[22,37]. As another advantage, the present $Bi_2Te_2Se$ films are sufficiently thin to permit tuning the Fermi level position over a wide range by applying a back gate voltage.

$Bi_2Te_2Se$ forms rhombohedral (space group $R\bar{3}m$) crystals consisting of hexagonally close-packed atomic layers of five atoms (quintuple layer) arranged along the c-axis as follows: $Se^{(1)}/Te^{(1)} - Bi - Se^{(2)}/Te^{(2)} - Bi - Se^{(1)}/Te^{(1)}$ (see Figure 1a), with the lattice constants a = 4.283 Å and c = 29.846 Å[24]. Hexagonal boron nitride (hBN), which likewise forms a layered crystal with hexagonal symmetry, has a lattice constant of 1.45 Å, close to that of graphene. The alignment of $Bi_2Te_2Se$ on hBN proposed in Figure 1b is characterized by a small lattice mismatch of 1.5%. In addition, the layered crystal structure of $Bi_2Te_2Se$ and hBN promotes epitaxial growth by van der Waals epitaxy. Herein, the weak van der Waals interaction between the two materials relaxes the lattice-mismatch condition[25].

Thin $Bi_2Te_2Se$ films were grown by a catalyst free vapor solid method[27,37], using a horizontal tube furnace equipped with a quartz tube (tube diameter 2.5 cm). First, ultrapure $Bi_2Se_3$ and $Bi_2Te_3$ crystalline powders (Sigma Aldrich, 99.999%) were placed in the hot zone of the furnace. The hBN growth substrates were then prepared by micromechanical cleavage of hBN powder on $Si/SiO_x$ substrates, and mounted ~15 cm away from the hot zone within the colder downstream region. The tube was evacuated repeatedly down to a pressure of $p < 1$ mbar and flushed with ultrapure Argon to remove oxygen, followed by adjusting the Argon flow to 150 sccm and the pressure in the tube to 80 mbar. Subsequently, the furnace was heated to approximately 590°C and held at his temperature for 30 s - 6 min, followed by natural cool down without gas flow. Morphology and thickness of the films were found to sensitively depend on the position of the growth substrates and the growth time. The accessible thickness was determined to range between 10 and 500 nm.

As exemplified by the AFM images in Figure 1c and Figure 2b, as well as the SEM image in Figure 2a, $Bi_2Te_2Se$ grows on top of the BN in the form of regular platelets with hexagonal symmetry. The $Bi_2Te_2Se$ growth mechanism proposed in Figure 1b is supported by the observation that the crystal facets in all platelets are oriented parallel or tilted by 60° or 120° degree (see Figure 2c). In addition, it can be seen in Figure 2a that the nanoplatelets preferably grow parallel to one of the sharp edges of the hBN sheet that are formed during exfoliation. Since the edges of the hBN flake enclose an angle of precisely 30° they can be assigned to armchair and zigzag edges. Assuming that the edges marked in blue in Figure 2a are the armchair edges of the BN flakes, the growth mechanism suggested in Figure 1b gains further support (if these edges were the zigzag edges instead, the resulting misfit would be unrealistically high, specifically 26% in case that the $Bi_2Te_2Se$ flakes would be compressed, and 47% if they would be expanded).

Evidence for the formation of $Bi_2Te_2Se$ could be gained by Raman spectroscopy. The chalcogenides $Bi_2Se_3$, $Bi_2Te_3$ and their alloys $Bi_2(Se_xTe_{1-x})_3$ are highly Raman active in the low wavenumber region[28]. The Raman spectrum in Figure 1d, acquired with a laser wavelength of 633 nm, displays three peaks at 105.0cm$^{-1}$, 139.1cm$^{-1}$ and 149.6cm$^{-1}$, respectively. They can be identified as the $E_{2g}$ and the $A^2_{1g}$ peak (which is split into two components) of $Bi_2Te_2Se$[28]. The additional peak at 1365.5 cm$^{-1}$ (see Figure 1d, inset) corresponds to the $E_{2g}$ mode of the underlying hBN flake[29]. It is noteworthy that besides $Bi_2Te_2Se$, also $Bi_2Se_3$ could be grown on hBN by van der Waals epitaxy (see Supplementary Information).

For electrical transport studies, standard e-beam lithography was performed to provide individual $Bi_2Te_2Se$ platelets with Ti(4 nm)/Au(200 nm) contacts in Hall-bar or van-der-Pauw geometry. Figure 3a displays the high field part of the Hall resistance $R_{xy}$ as a function of the magnetic field B at T = 1.5 K for a 45 nm thick $Bi_2Te_2Se$ platelet on a ~50 nm thick hBN sheet. The $R_{xy}$ signal clearly features SdH oscillations, as highlighted by the first derivative $dR_{xy}/dB$ in Figure 3a. In a prototypic two-dimensional (2D) electron gas, SdH oscillations are periodic in 1/B, with a periodicity given by[23]

$$2\pi n = A_F \frac{\hbar}{eB}, \qquad (1)$$

where n is the Landau level (LL) index, and $A_F$ is the external cross-section of the Fermi surface. For 2D electron systems, the formation of LLs depends only on the B-field component $B_\perp = B\cos(\theta)$ normal to the surface. Accordingly, if the observed quantum oscillations originated from the topological 2D surface states, the maxima/minima of the SdH oscillations should shift by $1/B\cos(\theta)$ upon tilting the sample in the magnetic field B. Indeed, in the plot of the amplitude of the SdH oscillations $\Delta R_{xy}$ (= $R_{xy}$ - $<R_{xy}>$, where $<R_{xy}>$ is a smoothed background) as a function of $B_\perp$ in Figure 3b, all curves coincide up to an angle of 50°. To further underscore the 2D character, the inset of Figure 3b demonstrates that the dependence of the position of the n=22 peak at B = 7.7 T (for θ=0°) on the tilting angle θ can be smoothly fitted with a 1/cos(θ) function.

Having confirmed the 2D nature of the SdH oscillations, we now address the position of the Fermi level in our samples. The Fermi energy can be calculated from

$$E_F = m_{cyc} v_F^2, \qquad (2)$$

where $m_{cyc}$ is the cyclotron mass and $v_F$ is the Fermi velocity. According to the Lifshitz-Kosevich (LK) theory[30] the thermal damping of the SdH oscillations provides a suitable estimate for the cyclotron mass $m_{cyc}$. From the plot of $\Delta R_{xy}$ vs. 1/B at different temperatures in Figure 4a, the oscillation amplitude is seen to decrease with increasing temperature. Figure 4b displays the amplitude of the n=16 peak at B = 11 T, normalized according to the LK formula:

$$\frac{\Delta \sigma_{xy}(T)}{\Delta \sigma_{xy}(0)} = \frac{\lambda(T)}{\sinh \lambda(T)}, \qquad (3)$$

where $\lambda(T) = 2\pi^2 k_B T m_{cyc}/(\hbar eB)$. Fitting of the data yields a value of $m_{cyc}/m_0$ = 0.15, which agrees well with previous reports on $Bi_2Te_2Se$ bulk crystals[20]. Furthermore, in order to access the Fermi vector $k_F$, we plot the position 1/B of the SdH maxima as a function of the LL index n (see Figure 5b below). The LL index is obtained by plotting the SdH maxima as a function of n+1/4 (since we are using the $R_{xy}$ signal, see Supplementary Information) and extrapolating to $1/B \to 0$ (i.e. $B \to \infty$). The intersection with the abscissa defines the first LL. By fitting the data with the Onsager relation (eq. (1)) and assuming a circular 2D Fermi surface with $A_F = \pi k_F^2$, one obtains a value of $k_F$ = 0.073 Å$^{-1}$. On this basis, a surface carrier density of $4.2 \cdot 10^{16} m^{-2}$ is calculated from $n_s = k_F^2 / 4\pi$. This value agrees well with the surface carrier density of $5.4 \cdot 10^{16}$ m$^{-2}$, as derived from Hall measurements (see Supplementary Information), thus further supporting the 2D nature of the SdH oscillations.

Toward determining the Fermi energy using eq. 2, we assume Dirac fermions with linear dispersion relation and a Fermi velocity of $v_F = \hbar k_F / m_{cyc} = 5.7 \cdot 10^5$ m/s, which is close to the value of $v_F = 4.9 \cdot 10^5$ m/s determined by ARPES[16]. On this basis, a Fermi energy of $E_F$ = 272 meV above the Dirac point is obtained. By comparison, a value of $E_F$ = 223 meV is derived for $k_F$ = 0.073 Å$^{-1}$ from the ARPES-derived band structure.[16] We have furthermore tested the possibility to tune the Fermi level in our samples through the action of a back gate. As can be discerned from Figure 5a, which displays the extracted SdH oscillations for different back gate voltages $V_G$, the period $\Delta 1/B$ rises with increasing $V_G$. In Figure 5b, the peak positions of the SdH oscillations are plotted as a function of the LL index n. As discussed above, the slopes directly yield values for the Fermi vector $k_F$ via eq. 1. The corresponding $E_F$ values can then be determined with the aid of the ARPES-derived band structure of Bi$_2$Te$_2$Se, see Figure 5d.[16] In this manner, the error associated with the assumptions made to calculate the cyclotron mass can be avoided (for comparison, Figure S5 shows $E_F$ values derived from the cyclotron mass). The sketch in Figure 5c evidences that $E_F$ (of the bottom surface which is influenced by the back gate, see arguments below) can be shifted by approximately 160 meV through the back gate voltage. Similar Fermi level shifts have been observed in case of considerably thicker hBN sheets (≈200 nm), in accordance with the estimated gate capacitances (see Supplementary Information). The application of a gate voltage is expected to cause bending of the bands at the interface between dielectric (hBN) and semiconductor (Bi$_2$Te$_2$Se), and thus to create a depletion layer. The thickness of the latter, which corresponds to the penetration depth of the electric field in the semiconductor, is of the order of only 10 nm.[32] Accordingly, the bulk and the top surface of the platelet should be only little affected by gating.

The fact that the SdH oscillations show only a single period for all gate voltages (see Fig. 4b), indicates the contribution of only one surface state. A similar conclusion has recently been drawn from the magnetotransport behavior of $Bi_2Te_2Se$ nanoplatelets on $Si/SiO_x$[37]. We attribute this finding to a deterioration of the surface state at the top surface due to long exposure to air, an explanation which is in accordance with previous charge transport studies on $Bi_2Se_3$.[12,14] Moreover, the results of low temperature Hall measurements performed at different back gate voltages (see Supplementary Information), are consistent with the n-doped character of our samples.[20,21,22]. It is furthermore noteworthy that the Hall mobilities agree well with the values extracted from the SdH oscillations for -60V < $V_g$ < 20V (see Supplementary Information).

We attribute the very high surface state carrier mobility in the range of 8.000 to 20.000 $cm^2$/Vs to the reduced defect density in the films grown by van der Waals epitaxy. This assertion gains support by comparing $Bi_2Te_2Se$ thin films grown on hBN or $Si/SiO_x$ substrates (see Supplementary Information). Specifically, an average surface state mobility of ~4900 $cm^2$/Vs was determined for $Bi_2Te_2Se$ on hBN, approximately three times larger than the value of 1600 $cm^2$/Vs obtained in case of the $Si/SiO_x$ substrates.

In summary, we have successfully grown the topological insulator $Bi_2Te_2Se$ on hBN in an oriented manner by van der Waals epitaxy. Thus obtained thin films show enhanced surface mobilities, which enable the study of gate-dependent quantum oscillations for the first time in this material. The unique combination of high mobility surface charge transport with an efficient tunability of the Fermi level provides a suitable basis for studying novel, spin-related charge transport phenomena in such devices.

**Figure captions**

**Figure 1.** Van-der-Waals epitaxy of $Bi_2Te_2Se$ on hBN. Schematic illustration of (a) the crystal structure of $Bi_2Te_2Se$, and (b) the proposed epitaxial growth mode on hBN. (c) Representative AFM image of $Bi_2Te_2Se$ nanoplatelets grown on hBN. All platelets are oriented parallel or rotated by an angle of 120° (indicated by the arrows). (d) Raman spectrum of $Bi_2Te_2Se$ on hBN ($\lambda_{exc}$ = 633 nm). The low wavenumber region can be fitted (green curve) using three Lorentzians (blue curves). The inset amplifies the wavenumber region 1300-1450 $cm^{-1}$, with a Lorentzian fit to the data (green curve).

**Figure 2.** (a) SEM image of $Bi_2Te_2Se$ nanoplatelets grown on hBN. The armchair and zigzag edges of the hBN sheet are marked in blue and red, respectively. (b) AFM image of the area marked in (a). (c) Histogram of the orientation distribution of the $Bi_2Te_2Se$ nanoplatelets grown on hBN.

**Figure 3.** (a) High-field Hall resistance (black curve) and its first derivative (green curve) as a function of the magnetic field showing pronounced Shubnikov-de-Haas (SdH) oscillations. (b) Amplitude of the SdH oscillations as a function of $1/(B\cos\theta)$ for different tilting angles $\theta$ (between surface normal and magnetic field). The inset shows the position $1/B$ of the SdH peak with the LL index n=22 (arrow) for different angles $\theta$, combined with a $1/\cos\theta$ fit (green curve) to the data.

**Figure 4.** (a) Amplitude of the SdH oscillations as a function of 1/B for different temperatures and (b) the normalized amplitude of the SdH maximum with the LL index n=15 (arrow) as a function of temperature. The data fit (solid line) was obtained using the Lifshitz-Kosevich theory (see main text).

**Figure 5.** (a) Amplitude of the SdH oscillations as a function of 1/B for different back gate voltages. (b) Plot of LL index 1/B versus number of maximum for different gate voltages. The data fits (solid lines) were performed using Eq.1. (c) Sketch of the position of the Fermi level of the bottom surface in the $Bi_2Te_2Se$ samples for different back gate voltages. (d) Calculated Fermi energy as a function of back gate voltage.

**Figure 1**

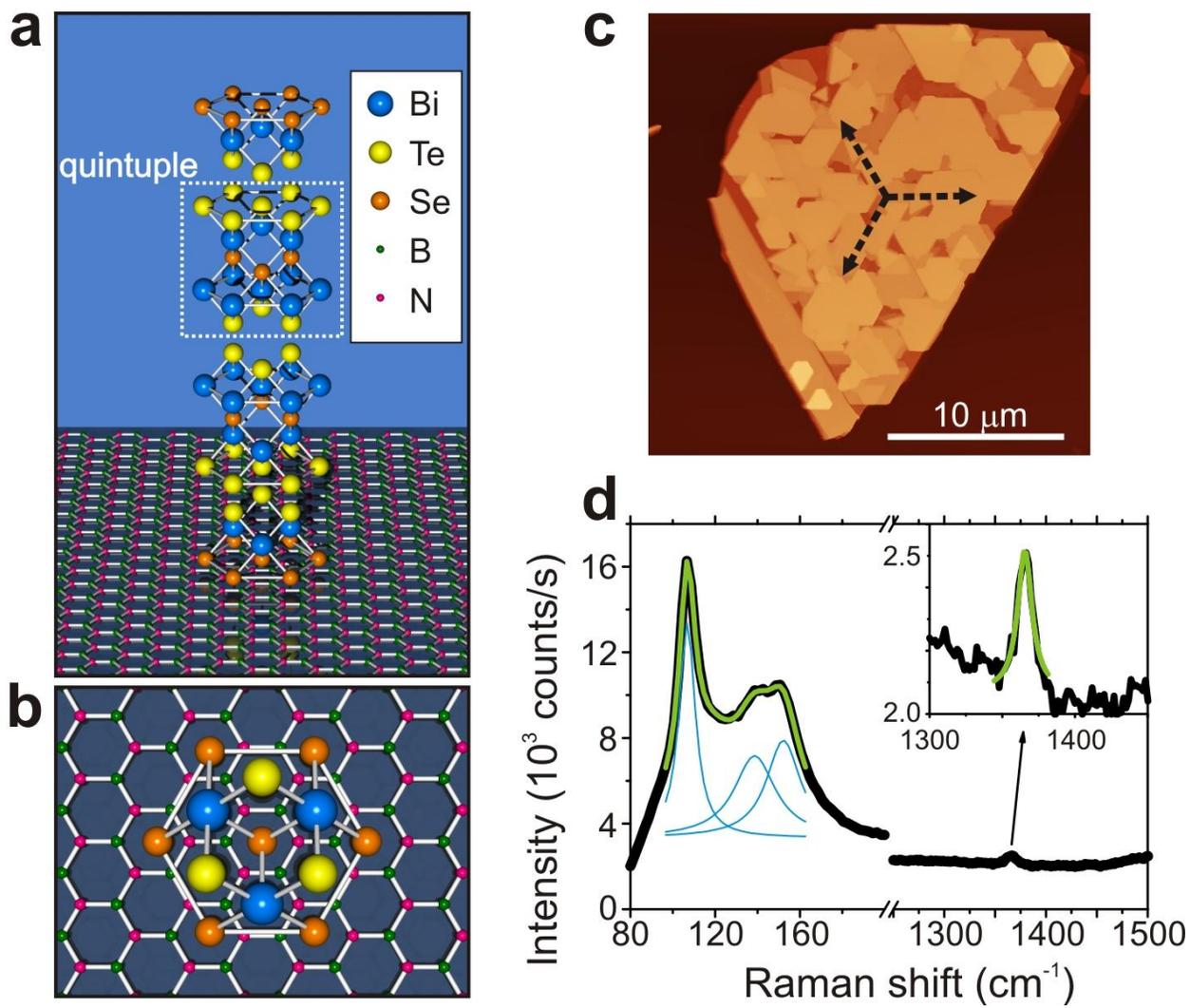

**Figure 2**

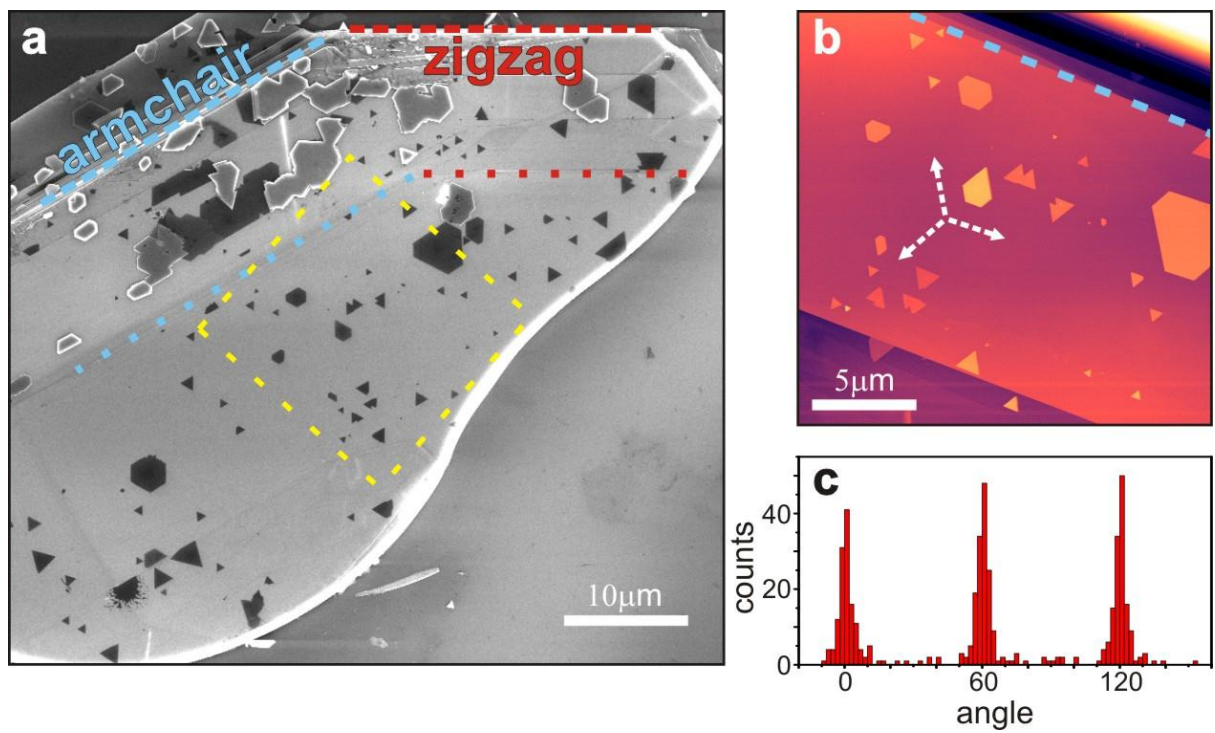

**Figure 3**

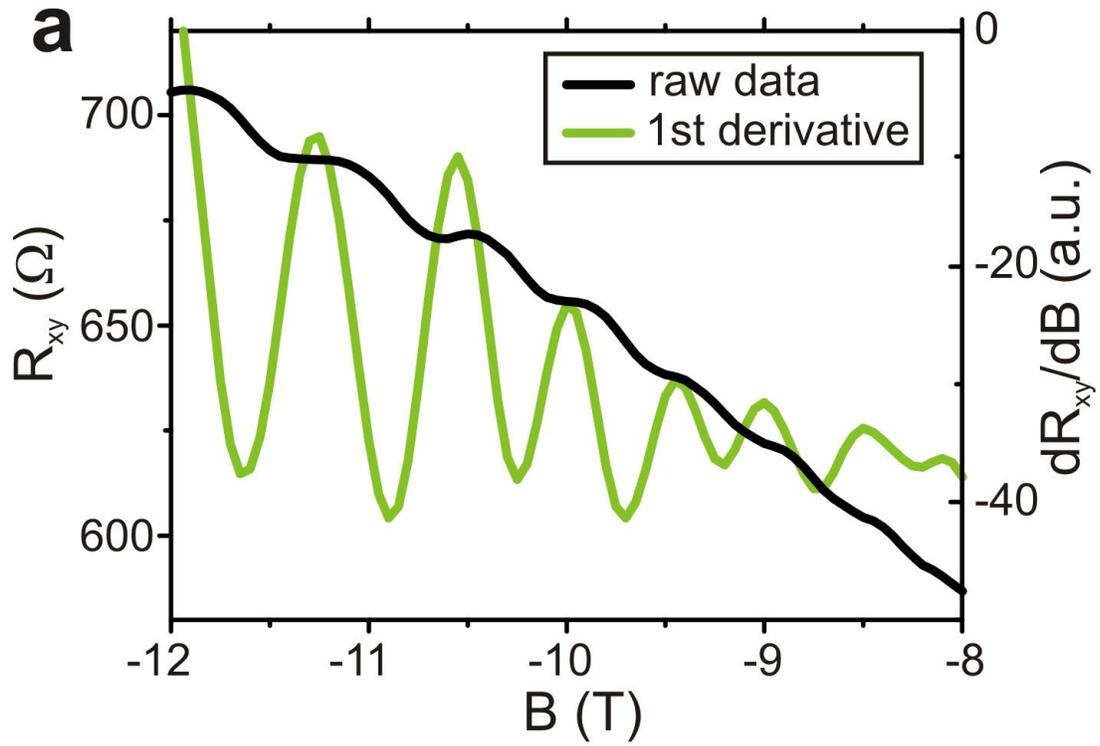

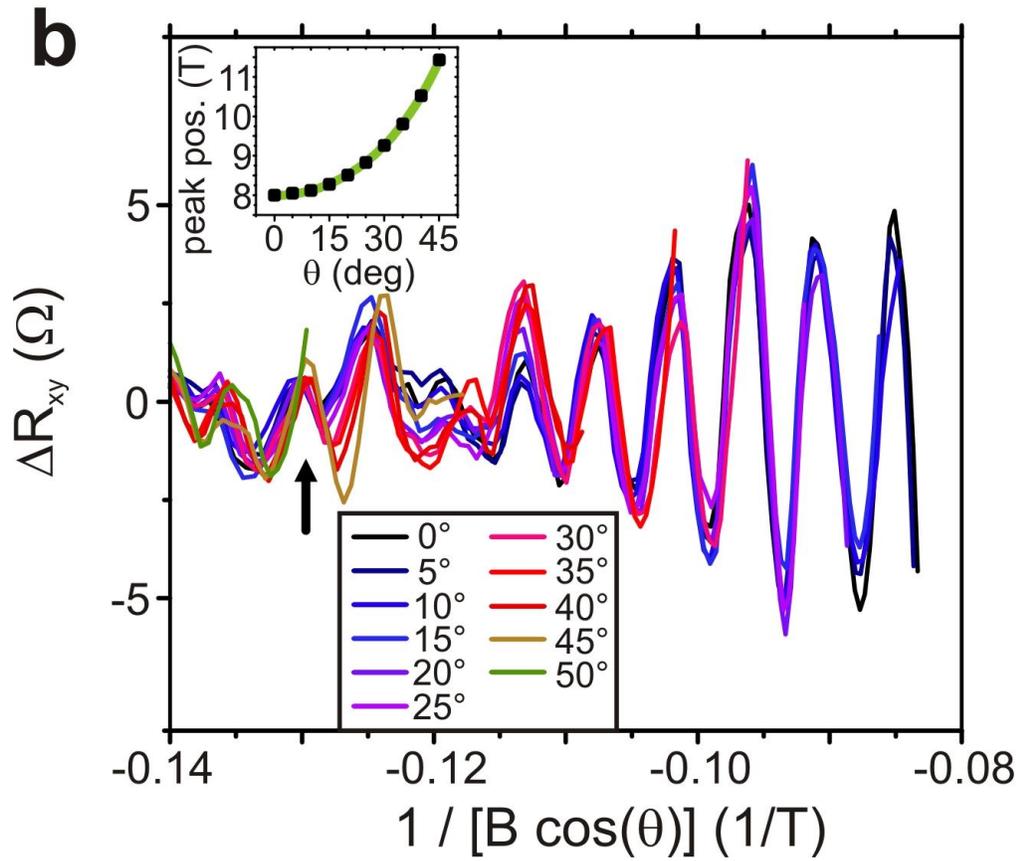

**Figure 4**

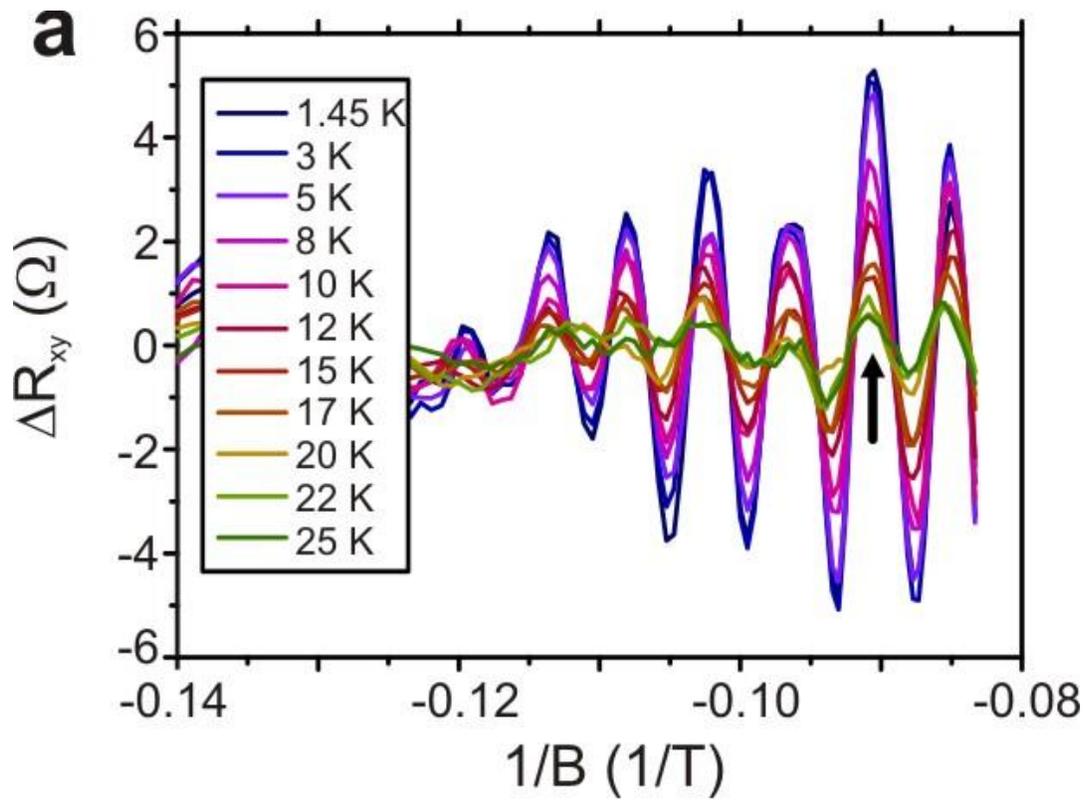

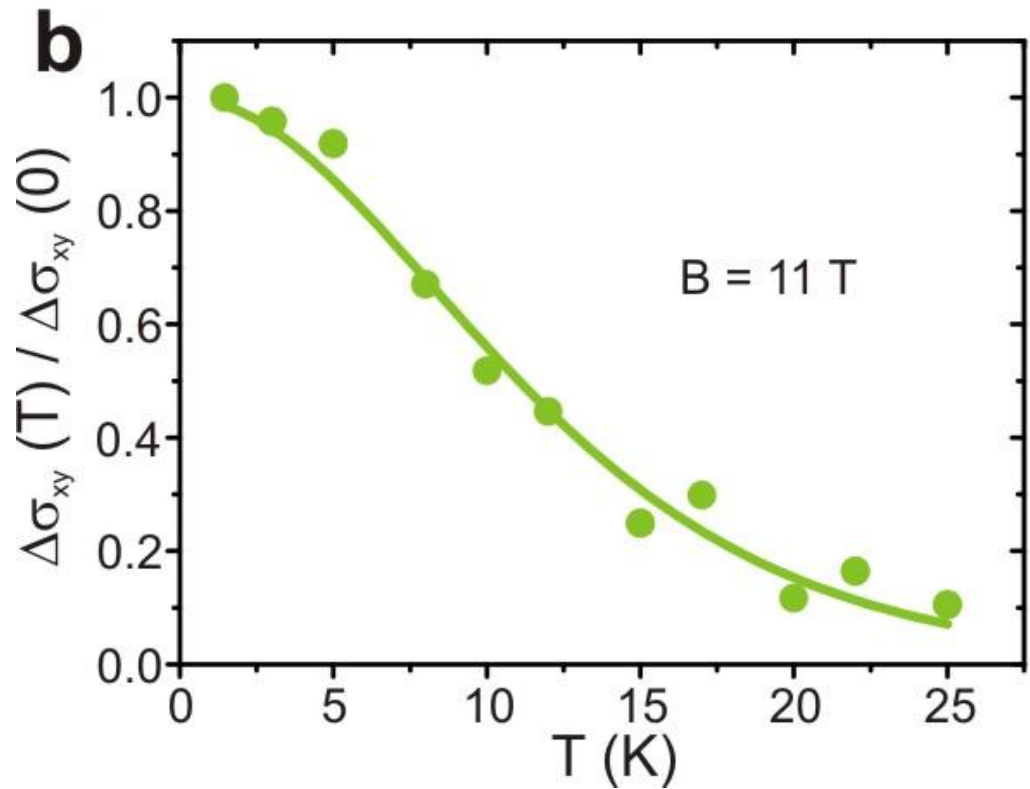

**Figure 5**

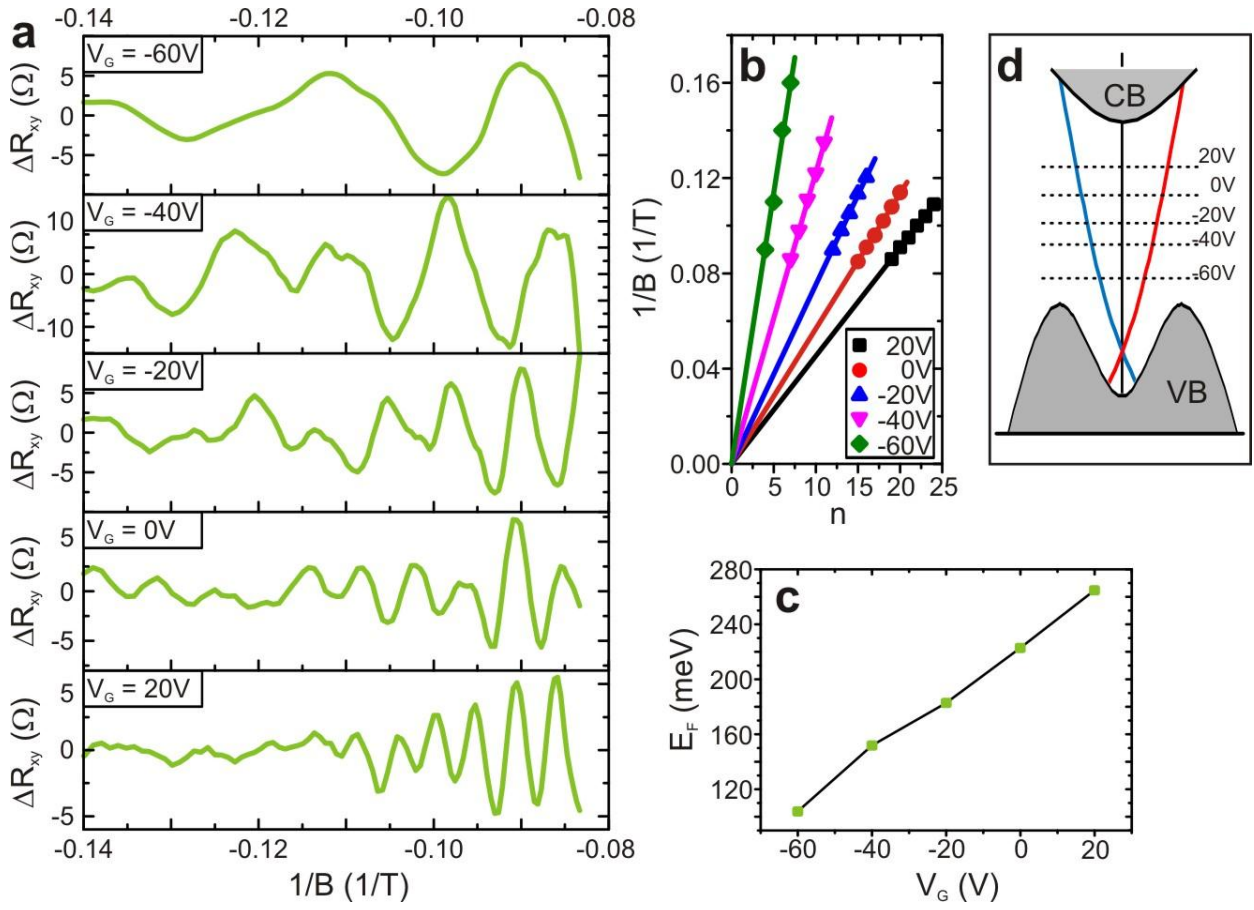

## ASSOCIATED CONTENT

**Supporting Information**.

Details on preparation of hBN flakes; Additional AFM and Raman data for $Bi_2Te_2Se$ films on hBN; Growth of $Bi_2Se_3$ on hBN; Hall data for the device discussed in the main text and another device; Calculation of the surface mobility from SdH data; Estimation of gate capacitance; Comparison of Fermi energy, as derived from ARPES band structure and SdH oscillations; Model for surface state mobility as a function of carrier density; Comparison between $R_{xx}$ and $R_{xy}$; Hall data for a 12nm thin sample. This material is available free of charge via the Internet at http://pubs.acs.org.

## AUTHOR INFORMATION

**Corresponding Author**

*P.G.: E-mail: p.gehring@fkf.mpg.de

**Present Addresses**

§Shanghai Institue of Microsystem and Information Technology, Chinese Academy of Sciences, Shanghai 200050, China

## ACKNOWLEDGMENT

The authors are grateful to Prof. Xiaoming Xie for providing the hBN powder and to E. C. Peters for help with the micromechanical cleavage of hBN. B. Krauss is acknowledged for support with the Raman set-up.

TOC

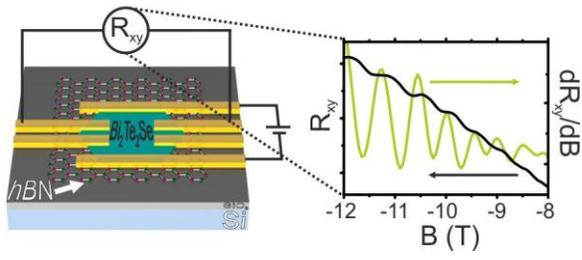